\documentstyle[12pt,epsf]{ioplppt} 
\begin{document} 
 
\title{ Shearing the Vacuum - Quantum Friction } 
\author{JB Pendry} 
 
\begin{abstract} 
We consider two perfectly smooth featureless surfaces at T=0, defined only by their  
respective dielectric functions, separated by a finite distance, and ask the question  
whether they can experience any friction when sheared parallel to their interface. We  
find large frictional effects comparable to everyday frictional forces provided that the  
materials have resistivities of the order of 1 m-W and that the surfaces are in close  
proximity. The friction depends solely on the reflection coefficients of the surfaces to  
electromagnetic waves and its detailed behaviour with shear velocity and separation is  
dictated by the dispersion of the reflectivity with frequency. 
\end{abstract} 
 
\address{ The Blackett Laboratory, 
Imperial College, 
London SW7 2BZ, 
UK} 
 
%
%
\maketitle

\section{Introduction} 
 
Friction is a problem of great practical importance, usually associated with complex  
systems that are difficult to characterise. There has been much discussion of the  
mechanisms lying behind friction [1,2] but despite many profound insights no all-  
embracing theory has been formulated to date and it is not the objective of this paper  
to propose one. Instead our aim is to focus on the simplest possible system, bereft of  
every complexity, and ask how friction arises. For simplicity we restrict ourselves to  
T=0. 
 
We consider two perfectly smooth featureless surfaces, parallel but not in contact,  
defined only by their electromagnetic reflection coefficients. We now shear the  
surfaces with relative velocity v and calculate the friction: see figure 1. Note the  
absence any roughness, a quality normally associated with friction. How is one  
surface `aware' of the other's motion? We can detect relative motion of a dielectric  
surface: reflect an electromagnetic wave from the surface and the reflection coefficient  
will show an asymmetry along the direction of motion, see figure 2. 
 
The two incident waves experience opposite Doppler shifts in the reference frame of  
the dielectric and, assuming dispersion with frequency, the two waves experience  
different reflection coefficients. If the surfaces are hot they will naturally emit  
radiation, but even a cold surface will be surrounded by a radiation field due to zero  
point quantum fluctuations. This exchange of virtual photons is the frictional  
mechanism we study in this paper. We shall show, contrary to a natural suspicion that  
such an effect should be extremely small, that the effect is large: comparable to other  
contributions to friction. 
 
Our point of view is that friction is the exchange of momentum between two surfaces  
and therefore we must be able to describe this process as an exchange of particles  
since all forces are ultimately mediated by a particle. In the case of friction with wear,  
atoms are exchanged between the surfaces either singly or in larger fragments. This is  
the classic frictional mechanism to which we can all relate: the grating of two rough  
surfaces in intimate contact. But it is only one mechanism, and it is understood that  
friction can occur even when there is negligible wear. Other means of momentum  
exchange are possible. 
 
Another particle that can be exchanged is the electron. This need not imply charging  
of the surfaces provided that equal numbers flow in opposite directions. Obviously  
metals, where the electrons are freely moving, are the prime candidates. Since electron  
density decreases exponentially outside a surface this mechanism will be short ranged  
limited to a scale dictated by the work function. We shall give brief consideration to  
this mechanism and show that it creates forces comparable to our electromagnetic  
mechanism. 
 
Finally there is the photon. This may mediate the electrostatic forces operating  
between charge distributions on opposite surfaces, or it may play a more subtle role  
involving the zero point fluctuations. Forces involving photons will be long-ranged  
because there is no work function preventing the photon's escape from a surface. It is  
with the photon that we shall concern ourselves here. We do not seek a universal  
explanation of friction. Our aim is far more modest: only to demonstrate the simplest  
possible model that exhibits friction by exchange of photons. 
 
Frictional forces on moving charges outside dielectrics have been studied for many  
years in the context of electron microscopy and nuclear radiation [3,4,5,6,7,8,9] and  
are well understood. The system with which we are concerned is different: the two  
surfaces we shear against one another are assumed to be locally electrically neutral.  
Quantum mechanically the dielectrics will experience internal charge fluctuations, and  
images of these charges in the other dielectric will create forces. We are already  
familiar with the force normal to the interface: the Van der Waals force, but there is in  
addition a parallel component because the image will lag slightly behind the charge  
creating it. We find that both forces can be calculated rather simply in terms of the  
reflection coefficients of the two surfaces. Furthermore their magnitude is not small  
provided that the surfaces are in close contact as in a normal friction experiment, and  
provided that the surfaces have a resistivity of the order of 1 m-W. This latter  
condition relates to the density of electromagnetic states available for dissipating  
energy. Immediately outside a surface this is proportional to, 
\begin{equation} 
\Im R\left( \omega  \right) 
\end{equation} 
where $R\left( \omega  \right)$ is the reflection coefficient of the surface. For a purely  
resistive sample this reduces to, 
\begin{equation} 
\Im {{i{\sigma  \over {\omega \varepsilon _0}}} \over {2+i{\sigma  \over {\omega  
\varepsilon _0}}}} 
\end{equation} 
where $\sigma $ is the conductivity. The effect is that of a washboard in which  
excitations on opposite surfaces with the same wave vector, $k$, grate against one  
another with frequency, 
\begin{equation} 
\omega =kv 
\end{equation} 
Given that the longest wave vectors generally correspond to wavelengths no less than  
an atomic spacing or so, typical velocities of $1{\rm ms}^{-1}$ will result in  
frequencies in the low GHz range. Adjusting the conductivity, $\sigma $, to maximise  
equation (2) 
\begin{equation} 
\sigma \approx 1\left( {m\Omega } \right)^{-1} 
\end{equation} 
To exchange momentum between the surfaces we need to create an excitation on each  
of the surfaces with equal and opposite momentum, not necessarily with equal and  
opposite energy though the total energy must sum to, 
\begin{equation} 
E=\hbar \omega =\hbar kv 
\end{equation} 
Thus energy radiates from the interface in correlated pairs of excitations. 
 
The possibility of radiating energy from dielectrics in motion is not a new one.  
Perhaps the most celebrated antecedent is the accelerated mirror concept: a mirror  
accelerated in the vacuum can be expected to produce electromagnetic radiation  
through its interaction with vacuum fluctuations [10]. More recently it has been  
suggested that sonoluminescence can be due to the acceleration of a dielectric fluid  
[11,12], though there is still some controversy about whether there is enough  
acceleration to give the observed effect [13]. All these phenomena are characterised  
by the extremely small amounts of radiation predicted and the difficulty of detecting  
it. Acceleration is a motion that cannot be removed by change of reference frame. A  
shear motion is another and we shall show that energy is also radiated in this situation.  
The big difference from the accelerated dielectric is that relatively large amounts of  
energy can be radiated under conditions of shear. 
 
The possibility of a dissipative component to Van der Waals forces has been  
considered in earlier work. In some instances the authors have been interested in other  
questions than frictional forces [14], and in other instances the formulae obtained  
differ from ours in crucial respects: for example the paper by Teodorovitch [15].  
Schaich and Harris [14] argue that that Teodorovitch is in error and that is our  
conclusion too. Levitov [16] presents some calculations of similar quantities, but  
without giving details, and his conclusions differ in some important aspects from  
those presented here: in particular his estimate of the forces is very small compared to  
ours. The work closest to ours is that by Annett and Echenique [17,18] on the friction  
experience by a neutral atom above a surface. Liebsch [19] also considered frictional  
forces on neutral atoms at surfaces. In this paper we give for the first time a simple  
and general derivation of the frictional forces between dielectrics. We draw attention  
for the first time to the important fact that if this frictional mechanism is to produce  
large effects, the relevant electromagnetic density of states must be maximised. This  
means choosing a resistivity of the order of 1 m-W as discussed above. 
 
We have eschewed complex diagrammatic formulations of this simple problem. They  
obscure the clarity of the situation and are prone to user error when the system is time  
dependent. Instead we give two derivations. The first given in section 2 is an intuitive  
one that produces the main results quickly and in some generality. We re-derive the  
same result in sections 3 and 4 by considering a simple model of a dielectric which  
dissipates energy through a set of harmonic oscillators, solving that model using a  
Lagrangian formulation to construct a quantum mechanical equation of motion for the  
harmonic oscillators. The latter method sheds light on the quantum mechanical  
processes at work, and highlights the two photon nature of the frictional process. 
 
By way of comparison we analyse in section 5 the frictional forces generated by  
exchange of electrons and show that these forces are of similar magnitude to photon  
based forces. Finally we discuss the possibility that light may be emitted as part of the  
frictional process and show that this possibility does not occur within our simple  
model, unless the dielectrics are sheared at unreasonably large velocities. 
 
\section{ Poor Man's Derivation of Quantum Friction } 
 
The final result for quantum friction is a simple one, and leads us to suspect that we  
could find a simple way of deriving it. It turns out that this is indeed that case, and in  
the process provides an interesting link with the conventional Van der Waals force  
between two surfaces. 
 
Imagine that we have a wave incident on a surface, 
\begin{equation} 
A\hat {\bf K}_p^+\exp \left( {ik_xx+ik_yy+iK_zz} \right) 
\end{equation} 
where the polarisation is chosen to be {\it p-}type and we work in the electrostatic  
limit neglecting the velocity of light, 
\begin{equation} 
\hat {\bf K}_p^\pm ={{c_0} \over \omega }\left[ {\matrix{{k_x}&{k_y}&{K_z=\pm  
i\sqrt {k_x^2+k_y^2}}\cr 
}} \right] 
\end{equation} 
In this limit the contribution of the {\it s-}polarised state is negligible. Reflection from  
the surface results in a total wavefield of, 
\begin{equation} 
\eqalign{&A\hat {\bf K}_p^+\exp \left( {ik_xx+ik_yy+iK_zz} \right)\cr 
  &+AR_{1pp}\left( {\omega +k_xv} \right)\hat {\bf K}_p^-\exp \left( {ik_xx+ik_yy- 
iK_zz} \right)\cr} 
\end{equation} 
Since we know the total wavefield we can calculate the force from the Maxwell stress  
tensor in vacuo, 
\begin{equation} 
T_{ij}={\textstyle{1 \over 2}}\left\{ \matrix{+\varepsilon  
_0E_i^{}E_j^*+\varepsilon _0E_i^*E_j^{}-\varepsilon _0\delta _{ij}{\bf E}\cdot  
{\bf E}^*\hfill\cr 
  +\mu _0H_i^{}H_j^*+\mu _0H_i^*H_j^{}-\mu _0\delta _{ij}{\bf H}\cdot {\bf  
H}^*\hfill\cr} \right\} 
\end{equation} 
The {\it H-}field being parallel to the surface makes no contribution to the force  
acting across the plane of the surface, but the electric fields give, 
\begin{equation} 
\eqalign{&F_x=2\left| A \right|^2{{\varepsilon _0c_0^2} \over {\omega ^2}}k_x\sqrt  
{k_x^2+k_y^2}\Im R_{1pp}^{}\left( {\omega +k_xv} \right)\cr 
  &F_z=2\left| A \right|^2{{\varepsilon _0c_0^2} \over {\omega ^2}}\left(  
{k_x^2+k_y^2} \right)\Re R_{1pp}^{}\left( {\omega +k_xv} \right)\cr} 
\end{equation} 
Note how the $x-$component of the force vanishes unless there is some dissipation in  
the system.  
 
Next we ask what might be the source of the incident wavefield? If a second surface is  
brought close to the first surface there will be a wavefield outside this second surface  
whose amplitude is given at $T=0$ by, 
\begin{equation} 
\left| A \right|^2={{\hbar \left| \omega  \right|} \over {2\varepsilon _0}}{{dN\left(  
{{\bf k},\omega } \right)} \over {d\omega }}d\omega  
\end{equation} 
where ${{dN\left( {{\bf k},\omega } \right)} \mathord{\left/ {\vphantom {{dN\left(  
{k,\omega } \right)} {d\omega }}} \right. \kern-\nulldelimiterspace} {d\omega }}$ is  
the density of electromagnetic states outside the second surface and the prefactor of  
${{\hbar \left| \omega  \right|} \mathord{\left/ {\vphantom {{\hbar \left| \omega   
\right|} {2\varepsilon _0}}} \right. \kern-\nulldelimiterspace} {2\varepsilon _0}}$  
normalises the wavefield to a total energy of ${{\hbar \left| \omega  \right|}  
\mathord{\left/ {\vphantom {{\hbar \left| \omega  \right|} 2}} \right. \kern- 
\nulldelimiterspace} 2}$ per mode. Further we can express the density of states at a  
distance $d$ from the second surface in terms of the reflection coefficient of the  
surface, 
\begin{equation} 
{{dN\left( {{\bf k},\omega } \right)} \over {d\omega }}={{\exp \left( {-2kd} \right)}  
\over {2\pi c_0^2k}}\omega \Im R_{2pp}^{}\left( \omega  \right) 
\end{equation} 
where we have assumed that the second surface is stationary in our frame of reference.  
Hence on substituting, 
\begin{equation} 
\left| A \right|^2={{\hbar \left| \omega  \right|} \over {2\varepsilon _0}}{{\exp \left(  
{-2kd} \right)} \over {2\pi c_0^2k}}\omega \Im R_{2pp}^{}\left( \omega   
\right)d\omega  
\end{equation} 
gathering together terms and integrating the forces over all frequencies and momenta  
parallel to the surface, 
\begin{equation} 
\eqalign{&F_x=4\sum\limits_{k_x,k_y}^{} {}\int_{-\infty }^{+\infty } {{{\hbar  
\left| \omega  \right|} \over {2\varepsilon _0}}}{{\exp \left( {-2kd} \right)} \over  
{2\pi c_0^2}}{{\varepsilon _0c_0^2} \over \omega }k_x\Im R_{1pp}^{}\left(  
{\omega +k_xv} \right)\Im R_{2pp}^{}\left( \omega  \right)d\omega \cr 
  &F_z=2\sum\limits_{k_x,k_y}^{} {}\int_{-\infty }^{+\infty } {{{\hbar \left|  
\omega  \right|} \over {2\varepsilon _0}}}{{\exp \left( {-2kd} \right)} \over {2\pi  
c_0^2}}{{\varepsilon _0c_0^2} \over \omega }k\Im \left[ {R_{1pp}^{}\left( {\omega  
+k_xv} \right)R_{2pp}^{}\left( \omega  \right)} \right]d\omega \cr} 
\end{equation} 
We have also added a factor of two because of a symmetrical process in which zero  
point waves emitted from surface one exert a force on surface two. Note the similarity  
between these two forces. The first, representing the frictional force between the two  
surfaces, vanishes when $v=0$ because of a symmetrical summation over an  
antisymmetrical function of $k_x$. The second represents the conventional Van der  
Waals force as, for example, derived in [20] and remains finite when $v=0$.  
 
Had we wished to study the effect of temperature we could have modified (11) to  
include the thermal contribution to radiation emitted from a free surface. 
 
We can simplify the expression for the frictional force as follows. First we transform  
the summation over $\bf k$ to an integral, 
\begin{eqnarray} 
\matrix{F_x&=&{\hbar  \over \pi }{{L^2} \over {\left( {2\pi } \right)^2}}\int_{- 
\infty }^{+\infty } {k_xdk_x}\int_{-\infty }^{+\infty } {\exp \left( {-2kd}  
\right)dk_y}\hfill\cr 
 & &\times \int_{-\infty }^{+\infty } {\Im R_{1pp}^{}\left( {\omega +k_xv}  
\right)\Im R_{2pp}^{}\left( \omega  \right) {\rm sgn} \left( \omega  \right)d\omega  
}\hfill\cr 
  &=&{\hbar  \over \pi }{{L^2} \over {\left( {2\pi } \right)^2}}\int_0^{+\infty }  
{k_xdk_x}\int_{-\infty }^{+\infty } {\exp \left( {-2kd} \right)dk_y}\hfill\cr 
  &&\times \int_{-\infty }^{+\infty } {\left[ {\Im R_{1pp}^{}\left( {\omega +k_xv}  
\right)-\Im R_{1pp}^{}\left( {\omega -k_xv} \right)} \right]\Im R_{2pp}^{}\left(  
\omega  \right) {\rm sgn} \left( \omega  \right)d\omega }\hfill\cr} 
\end{eqnarray} 
where we have exploited the symmetry of the reflection coefficient which follows  
from general principles of causality, 
\begin{equation} 
\Im R\left( {-\omega } \right)=-\Im R\left( {+\omega } \right) 
\end{equation} 
Next we make a similar rearrangement of the frequency integration to give, 
\begin{eqnarray} 
\matrix{F_x=&{\hbar  \over \pi }{{L^2} \over {\left( {2\pi }  
\right)^2}}\int_0^{+\infty } {k_xdk_x}\int_{-\infty }^{+\infty } {\exp \left( {-2kd}  
\right)dk_y}\hfill\cr 
  &\times 2\int_0^{+\infty } {\left[ {\Im R_{1pp}^{}\left( {\omega +k_xv} \right)- 
\Im R_{1pp}^{}\left( {\omega -k_xv} \right)} \right]\Im R_{2pp}^{}\left( \omega   
\right)d\omega }\hfill\cr} 
\end{eqnarray} 
One final manipulation gives, 
\begin{eqnarray} 
\matrix{F_x&=&{\hbar  \over \pi }{{L^2} \over {\left( {2\pi }  
\right)^2}}\int_0^{+\infty } {k_xdk_x}\int_{-\infty }^{+\infty } {\exp \left( {-2kd}  
\right)dk_y}\hfill\cr 
 & &\times 2\left[ \matrix{+\int_0^{+\infty } {\Im R_{1pp}^{}\left( {\omega +k_xv}  
\right)\Im R_{2pp}^{}\left( \omega  \right)d\omega }\hfill\cr 
  -\int_0^{+\infty } {\Im R_{1pp}^{}\left( {\omega -k_xv} \right)\Im  
R_{2pp}^{}\left( \omega  \right)d\omega }\hfill\cr} \right]\hfill\cr 
  &=&{\hbar  \over \pi }{{L^2} \over {\left( {2\pi } \right)^2}}\int_0^{+\infty }  
{k_xdk_x}\int_{-\infty }^{+\infty } {\exp \left( {-2kd} \right)dk_y}\hfill\cr 
&  &\times 2\left[ \matrix{+\int_{k_xv}^{+\infty } {\Im R_{1pp}^{}\left( \omega   
\right)\Im R_{2pp}^{}\left( {\omega -k_xv} \right)d\omega }\hfill\cr 
  -\int_0^{+\infty } {\Im R_{1pp}^{}\left( {\omega -k_xv} \right)\Im  
R_{2pp}^{}\left( \omega  \right)d\omega }\hfill\cr} \right]\hfill\cr} 
\end{eqnarray} 
Obviously we could have made the same manipulations with the role of the two  
surfaces inverted, 
\begin{eqnarray} 
\matrix{F_x=&{\hbar  \over \pi }{{L^2} \over {\left( {2\pi }  
\right)^2}}\int_0^{+\infty } {k_xdk_x}\int_{-\infty }^{+\infty } {\exp \left( {-2kd}  
\right)dk_y}\hfill\cr 
  &\times 2\int_0^{+\infty } {\left[ {\Im R_{1pp}^{}\left( {\omega +k_xv} \right)- 
\Im R_{1pp}^{}\left( {\omega -k_xv} \right)} \right]\Im R_{2pp}^{}\left( \omega   
\right)d\omega }\hfill\cr 
  F_x=&{\hbar  \over \pi }{{L^2} \over {\left( {2\pi } \right)^2}}\int_0^{+\infty }  
{k_xdk_x}\int_{-\infty }^{+\infty } {\exp \left( {-2kd} \right)dk_y}\hfill\cr 
  &\times 2\left[ \matrix{+\int_{k_xv}^{+\infty } {\Im R_{2pp}^{}\left( \omega   
\right)\Im R_{1pp}^{}\left( {\omega -k_xv} \right)d\omega }\hfill\cr 
  -\int_0^{+\infty } {\Im R_{2pp}^{}\left( {\omega -k_xv} \right)\Im  
R_{1pp}^{}\left( \omega  \right)d\omega }\hfill\cr} \right]\hfill\cr} 
\end{eqnarray} 
Adding the two expressions gives, 
\begin{eqnarray} 
\matrix{F_x=&4\hbar {{L^2} \over {\left( {2\pi } \right)^3}}\int_0^{+\infty }  
{k_xdk_x}\int_{-\infty }^{+\infty } {\exp \left( {-2kd} \right)dk_y}\hfill\cr 
  &\times \int_0^{k_xv} {\Im R_{2pp}^{}\left( \omega  \right)\Im R_{1pp}^{}\left(  
{k_xv-\omega } \right)d\omega }\hfill\cr} 
\end{eqnarray} 
It only remains to identify the reflection coefficients of a dielectric surface as, 
\begin{equation} 
\matrix{R_{ss}=+{{\cos \left( {\theta _i} \right)-\sqrt \varepsilon \cos \left( {\theta  
_r} \right)} \over {\cos \left( {\theta _i} \right)+\sqrt \varepsilon \cos \left( {\theta _r}  
\right)}},\;\;\;\;\;\hfill\cr 
  R_{pp}=-{{\cos \left( {\theta _r} \right)-\sqrt \varepsilon \cos \left( {\theta _i}  
\right)} \over {\cos \left( {\theta _r} \right)+\sqrt \varepsilon \cos \left( {\theta _i}  
\right)}}\hfill\cr} 
\end{equation} 
where the angles are now complex, 
\begin{equation} 
\eqalign{&\cos \left( {\theta _i} \right)={{K_z\left( {vac} \right)} \over {\left| {{\bf  
K}\left( {vac} \right)} \right|}}={{i\sqrt {k_x^2+k_y^2-\omega ^2c_0^{-2}}} \over  
{\omega \,c_0^{-1}}}\approx {{i\sqrt {k_x^2+k_y^2}} \over {\omega \,c_0^{- 
1}}},\cr 
  &\cos \left( {\theta _r} \right)={{K_z\left( \varepsilon  \right)} \over {\left| {{\bf  
K}\left( \varepsilon  \right)} \right|}}={{i\sqrt {k_x^2+k_y^2-\omega ^2\varepsilon  
c_0^{-2}}} \over {\omega \,\sqrt \varepsilon c_0^{-1}}}\approx {{i\sqrt  
{k_x^2+k_y^2}} \over {\omega \,\sqrt \varepsilon c_0^{-1}}}\cr} 
\end{equation} 
Hence, 
\begin{equation} 
\mathop {\lim }\limits_{k_x^2+k_y^2\to \infty }R_{ss}=0,\;\;\;\;\;\mathop {\lim  
}\limits_{k_x^2+k_y^2\to \infty }R_{pp}={{\varepsilon -1} \over {\varepsilon +1}} 
\end{equation} 
and, 
\begin{equation} 
\matrix{F_x=4\hbar {{L^2} \over {\left( {2\pi } \right)^3}}\int_0^{+\infty }  
{k_xdk_x}\int_{-\infty }^{+\infty } {\exp \left( {-2kd} \right)dk_y}\hfill\cr 
  \times \int_0^{k_xv} {\Im {{\varepsilon _1\left( \omega  \right)-1} \over  
{\varepsilon _1\left( \omega  \right)+1}}\Im {{\varepsilon _2\left( {k_xv-\omega }  
\right)-1} \over {\varepsilon _2\left( {k_xv-\omega } \right)+1}}d\omega }\hfill\cr} 
\end{equation} 
which is the same as our more hard-won expression of section 4. Note, however, that  
the formulation in terms of reflection coefficients is more general as it makes no  
assumption whatever about the internal structure of the surfaces. This may be  
important when considering systems in which the surface is intrinsically different  
from the bulk: surfaces with thin coatings being a pertinent example.  
 
Within this simple formulation it is easy to correct for higher order perturbations. We  
saw above how quantum fluctuations in a second surface result in a fluctuating  
wavefield incident on the first surface, amplitude $A$. However we can identify  
further contributions to the incident amplitude from waves that are reflected from the  
first surface, and then from the second surface to return again. The process can be  
repeated to give a corrected incident amplitude, 
\begin{eqnarray} 
\matrix{A^\prime &=A\left[ \matrix{1+e^{-2kd}R_{2pp}^{}\left( \omega   
\right)R_{1pp}^{}\left( {\omega +k_xv} \right)\hfill\cr 
  +\left\{ {e^{-2kd}R_{2pp}^{}\left( \omega  \right)R_{1pp}^{}\left( {\omega  
+k_xv} \right)} \right\}^2+\;\cdots \hfill\cr} \right]\hfill\cr 
  &=A\left[ {1-e^{-2kd}R_{2pp}^{}\left( \omega  \right)R_{1pp}^{}\left( {\omega  
+k_xv} \right)} \right]^{-1}\hfill\cr} 
\end{eqnarray} 
Hence correcting equation (14) for multiple scattering, 
\begin{eqnarray} 
\matrix{F_x=&4\sum\limits_{k_x,k_y}^{} {}\int_{-\infty }^{+\infty } {{{\hbar \left|  
\omega  \right|} \over {2\varepsilon _0}}}{{\exp \left( {-2kd} \right)} \over {2\pi  
c_0^2}}{{\varepsilon _0c_0^2} \over \omega }k_x\Im R_{1pp}^{}\left( {\omega  
+k_xv} \right)\Im R_{2pp}^{}\left( \omega  \right)\hfill\cr 
  &\times \left| {1-e^{-2kd}R_{2pp}^{}\left( \omega  \right)R_{1pp}^{}\left(  
{\omega +k_xv} \right)} \right|^{-2}d\omega \hfill\cr 
  F_z=&2\sum\limits_{k_x,k_y}^{} {}\int_{-\infty }^{+\infty } {{{\hbar \left|  
\omega  \right|} \over {2\varepsilon _0}}}{{\exp \left( {-2kd} \right)} \over {2\pi  
c_0^2}}{{\varepsilon _0c_0^2} \over \omega }k\Im \left[ {R_{1pp}^{}\left( {\omega  
+k_xv} \right)R_{2pp}^{}\left( \omega  \right)} \right]\hfill\cr 
  &\times \left| {1-e^{-2kd}R_{2pp}^{}\left( \omega  \right)R_{1pp}^{}\left(  
{\omega +k_xv} \right)} \right|^{-2}d\omega \hfill\cr} 
\end{eqnarray} 
In the examples we shall consider the surfaces are mainly resistive and multiple  
scattering corrections make only a qualitative change to our predictions even when the  
surfaces are close together. However when the surfaces support local modes,  
frequencies of these modes can be drastically shifted by proximity of a second surface  
strongly affecting their contribution to friction. 
 
\section{ A Classical Hamiltonian Description of Moving Surfaces } 
 
We have restricted ourselves to a system defined purely in terms of a classical  
macroscopic quantity: the dielectric constant, $\varepsilon \left( \omega  \right)$.  
Therefore it is appropriate that we begin by constructing the classical equations of  
motion of the system, before following the well worn path of quantisation via the  
Hamiltonian. We begin by constructing a Langrangian, then define momentum  
coordinates which are used to find the Hamiltonian. 
 
First consider a classical system comprising surface of a dielectric material in vacuo.  
We assume that the dielectric is a dynamic object defined by a continuum of harmonic  
oscillator modes. These modes are vital to our subsequent calculations because they  
will be responsible for transporting energy away from the surface in the frictional  
process. Although we shall define the modes through the losses they produce, that will  
also fix the real part of the dielectric function through causality as realised in the  
Kramers Kronig relationships. Since we assume that the surface is translationally  
invariant, each mode is defined partly by a wave vector parallel to the surface, $\bf k$,  
and partly by a second subscript, $j$, which may be associated with degrees of  
freedom normal to the surface, and is responsible for transportation of energy away  
from the surface. We need say nothing about the nature of the modes other than how  
they couple to the outside world and this we probe by a sheet of charge placed a  
distance $d$ above the surface. We can now write down the Lagrangian, 
\begin{equation} 
L_1=T-V=\sum\limits_{j{\bf k}}^{} {\dot s_{j{\bf k}}^2-\omega _{j{\bf  
k}}^2s_{j{\bf k}}^2-A_{\bf k}\beta _{j{\bf k}}^{}s_{j{\bf k}}^{}\exp \left( {-kd- 
i\Omega t} \right)} 
\end{equation} 
The first two terms on the right hand side define the harmonic oscillators. Although  
they are written for simplicity as a discrete summation over modes, we shall always  
take the continuum limit. The last term on the right hand side represents the coupling  
of each mode to the external charge. Note that it drops off exponentially with distance  
from the surface: at this stage we are mainly concerned with very short wavelength  
modes whose fields in the vacuum are largely electrostatic and therefore, 
\begin{equation} 
\matrix{k_x^2+k_y^2+k_z^2=0,\hfill\cr 
  k_z=i\sqrt {k_x^2+k_y^2}\hfill\cr} 
\end{equation} 
The point of introducing the external charge is to probe the electrical activity of the  
modes in the vacuum. Since for any frictional process we are interested in the mutual  
excitation of modes across the intervening vacuum, all relevant coupling must pass  
through the vacuum and is therefore probed by our test charge. We can calculate the  
coupling parameter, $\beta _{j{\bf k}}^{}$, by calculating $P$, the rate of dissipation  
of energy in the dielectric, in two ways then equating the results. 
 
First we use the Lagrangian equations of motion to calculate, 
\begin{equation} 
P_{\bf k}={{\Omega \,A_{\bf k}^2} \over {16}}\Im \sum\limits_j^{} {\,{{\beta  
_{j{\bf k}}^2\,} \over {\Omega ^2-\omega _{j{\bf k}}^2}}\;}={{\pi \,A_{\bf k}^2}  
\over {16}}\beta _{j{\bf k}}^2{{dN_{j{\bf k}}} \over {d\omega _{j{\bf k}}}} 
\end{equation} 
Note how it is essential that we take the continuum limit. Otherwise there is no  
contribution from the poles. 
 
Alternatively we may recognise that the test charge induces an image charge in the  
dielectric, 
\begin{equation} 
q^\prime=-\left[ {{{\varepsilon \left( \Omega  \right)-1} \over {\varepsilon \left(  
\Omega  \right)+1}}} \right]A_{\bf k}\exp \left( {ik_xx+ik_yy-i\Omega t} \right) 
\end{equation} 
and therefore the loss can be found from the rate of working of the test charge, 
\begin{equation} 
P^\prime_{\bf k} = {{A_{\bf k}^2k\Omega \varepsilon _0} \over 2}\Im \left[  
{{{\varepsilon \left( \Omega  \right)-1} \over {\varepsilon \left( \Omega  \right)+1}}}  
\right] 
\end{equation} 
where, 
\begin{equation} 
{{dN_{\bf k}} \over {d\omega _{\bf k}}} 
\end{equation} 
is the density of modes of wave vector $\bf k$. Equating $P$ and $P^\prime$ gives, 
\begin{equation} 
\beta _{j{\bf k}}^2{{dN_{j{\bf k}}} \over {d\omega _{j{\bf k}}}}={{8k\Omega  
\varepsilon _0} \over \pi }\Im \left[ {{{\varepsilon \left( \Omega  \right)-1} \over  
{\varepsilon \left( \Omega  \right)+1}}} \right] 
\end{equation} 
 
Next we write down the Lagrangian for two parallel surfaces separated by a distance  
$d$, for the moment assumed stationary with respect to one another: 
\begin{equation} 
\matrix{L_{12}=\sum\limits_{j{\bf k}}^{} {\dot s_{j{\bf k}1}^2-\omega _{j{\bf  
k}}^2s_{j{\bf k}1}^2}+\sum\limits_{j^\prime {\bf k}^\prime}^{} {\dot  
s_{j^\prime{\bf k}^\prime2}^2-\omega _{j^\prime {\bf k}^\prime }^2s_{j^\prime  
{\bf k}^\prime 2}^2}\hfill\cr 
  -\sum\limits_{jj^\prime}^{} {{{\beta _{j{\bf k}}^{}\beta _{j^\prime{\bf k}}^{}}  
\over {4k\varepsilon _0}}\exp \left( {-kd} \right)s_{j{\bf k}1}^{}s_{j^\prime{\bf  
k}2}^{}}\hfill\cr} 
\end{equation} 
We assume that the two surfaces  are identical. Note that the test charge has been  
removed and that a new term represents coupling between modes on opposite  
surfaces. Since the coupling is mediated by an electrostatic field in the vacuum, our  
previous calculation of the coupling parameter is valid here also.  
 
Finally the two surfaces are set in motion relative to one another as shown in figure 1.  
We shall assume that the relative velocity is small compared to the velocity of light.  
Choosing a frame of reference in which the surfaces have equal and opposite  
velocities, ${\textstyle{1 \over 2}}v$: 
\begin{eqnarray} 
\matrix{L_{12}\left( t \right)=&\sum\limits_{j{\bf k}}^{} {\dot s_{j{\bf k}1}^2- 
\omega _{j{\bf k}}^2s_{j{\bf k}1}^2}+\sum\limits_{j^\prime{\bf k}^\prime}^{}  
{\dot s_{j^\prime{\bf k}^\prime2}^2-\omega _{j^\prime{\bf  
k}^\prime}^2s_{j^\prime{\bf k}^\prime2}^2}\hfill\cr 
  &-\sum\limits_{jj^\prime}^{} {{{\beta _{j{\bf k}}^{}\beta _{j^\prime{\bf k}}^{}}  
\over {4k\varepsilon _0}}\exp \left( {-kd} \right)s_{j{\bf k}1}^{}s_{j^\prime{\bf  
k}2}^{}}\exp \left( {-ik_xvt} \right)\hfill\cr} 
\end{eqnarray} 
Only the coupling term changes: modes on opposite surfaces grate against one another  
in a washboard effect generating a frequency of $kv$. It is this finite frequency that  
will induce transitions in the system causing dissipation of energy. It is worth noting  
that the relevant frequencies range from zero up to a cut off which is imposed either  
by the exponential decay of the coupling, 
\begin{equation} 
\omega _{\max 1}=v/d 
\end{equation} 
or, if $d$ is very small, by the shortest wavelength fluctuations in the dielectric,  
usually of the order of $10^{-10}\rm m$. Thus the relevant frequencies are of the  
order of $10^{+10}{\rm Hz}$ for a shear velocity of $1{\rm ms}^{-1}$.  
 
The final step in the classical treatment is to extract an Hamiltonian from the  
Lagrangian. First we define canonical momenta, 
\begin{equation} 
t={{\partial L} \over {\partial \dot s}}=2\dot s 
\end{equation} 
which gives, 
\begin{eqnarray} 
\matrix{H_{12}\left( t \right)&=&\sum\limits_{j{\bf k}}^{} {t_{j{\bf k}1}^{}\dot  
s_{j{\bf k}1}^{}}+\sum\limits_{j^\prime{\bf k}^\prime}^{} {t_{j^\prime{\bf  
k}^\prime2}^{}\dot s_{j^\prime{\bf k}^\prime2}^{}}-L_{12}\left( t \right)\hfill\cr 
  &=&\sum\limits_{j{\bf k}}^{} {{\textstyle{1 \over 4}}t_{j{\bf k}1}^2+\omega  
_{j{\bf k}}^2s_{j{\bf k}1}^2}+\sum\limits_{j^\prime{\bf k}^\prime}^{}  
{{\textstyle{1 \over 4}}t_{j^\prime{\bf k}^\prime2}^2+\omega _{j^\prime{\bf  
k}^\prime}^2s_{j^\prime{\bf k}^\prime2}^2}\hfill\cr 
  &&+\sum\limits_{jj^\prime}^{} {{{\beta _{j{\bf k}}^{}\beta _{j^\prime{\bf  
k}}^{}} \over {4k\varepsilon _0}}\exp \left( {-kd} \right)s_{j{\bf  
k}1}^{}s_{j^\prime{\bf k}2}^{}}\exp \left( {-ik_xvt} \right)\hfill\cr} 
\end{eqnarray} 
 
\section{ A Quantum Description of Moving Surfaces } 
 
Now we introduce quantum mechanics into our classical picture using the  
conventional identification of, 
\begin{equation} 
t\to -i\hbar {\partial  \over {\partial s}} 
\end{equation} 
giving for the time dependent Schrödinger equation, 
\begin{eqnarray} 
\matrix{i{\partial  \over {\partial t}}\Psi \left( t \right)&=&H_{12}\left( t \right)\Psi  
\left( t \right)\hfill\cr 
  &=&+\sum\limits_{j{\bf k}}^{} {\left\{ {-{\textstyle{1 \over 4}}{{\partial ^2}  
\over {\partial ^2s_{j{\bf k}1}^{}}}+\omega _{j{\bf k}}^2s_{j{\bf k}1}^2}  
\right\}\Psi \left( t \right)}\hfill\cr 
  &&+\sum\limits_{j^\prime {\bf k}^\prime }^{} {\left\{ {-{\textstyle{1 \over  
4}}{{\partial ^2} \over {\partial ^2s_{j^\prime {\bf k}^\prime 2}^{}}}+\omega  
_{j^\prime {\bf k}^\prime }^2s_{j^\prime {\bf k}^\prime 2}^2} \right\}\Psi \left( t  
\right)}\hfill\cr 
  &&+\sum\limits_{jj^\prime }^{} {{{\beta _{j{\bf k}}^{}\beta _{j^\prime {\bf  
k}}^{}} \over {4k\varepsilon _0}}\exp \left( {-kd} \right)s_{j{\bf  
k}1}^{}s_{j^\prime {\bf k}2}^{}}\exp \left( {-ik_xvt} \right)\Psi \left( t  
\right)\hfill\cr} 
\end{eqnarray} 
Some general observations are in order. The final term in the Schrödinger equation is  
capable of creating excitations in the system of energy $\hbar kv$ as can be made  
apparent by defining annihilation and creation operators, 
\begin{equation} 
s_{j{\bf k}1}^\pm ={1 \over {\sqrt 2}}\left[ {-i{\partial  \over {\partial s_{j{\bf  
k}1}^{}}}\pm is_{j{\bf k}1}^{}} \right],\;\;\;\;\;s_{j{\bf k}2}^\pm ={1 \over {\sqrt  
2}}\left[ {-i{\partial  \over {\partial s_{j{\bf k}2}^{}}}\pm is_{j{\bf k}2}^{}} \right] 
\end{equation} 
hence 
\begin{equation} 
s_{j{\bf k}1}^{}s_{j^\prime {\bf k}2}^{}=-{1 \over 2}\left[ {s_{j{\bf k}1}^+- 
s_{j{\bf k}1}^-} \right]\left[ {s_{j^\prime {\bf k}2}^+-s_{j^\prime {\bf k}2}^-}  
\right] 
\end{equation} 
so that the interaction term becomes, 
\begin{eqnarray} 
&+\sum\limits_{jj^\prime }^{} {{{\beta _{j{\bf k}}^{}\beta _{j^\prime {\bf k}}^{}}  
\over {4k\varepsilon _0}}\exp \left( {-kd} \right)s_{j{\bf k}1}^{}s_{j^\prime {\bf  
k}2}^{}}\exp \left( {-ik_xvt} \right)\Psi \left( t \right)\cr 
  =&-{1 \over 2}\sum\limits_{jj^\prime }^{} {{{\beta _{j{\bf k}}^{}\beta _{j^\prime  
{\bf k}}^{}} \over {4k\varepsilon _0}}\exp \left( {-kd} \right)\left[ {s_{j{\bf k}1}^+- 
s_{j{\bf k}1}^-} \right]\left[ {s_{j^\prime {\bf k}2}^+-s_{j^\prime {\bf k}2}^-}  
\right]}\exp \left( {-ik_xvt} \right)\Psi \left( t \right)\cr 
& 
\end{eqnarray} 
A system originally in the ground state can only absorb energy that is shared between  
the two surfaces because an excitation is created on each surface. Frictional energy  
will be emitted from the interface in correlated pairs of excitations. 
 
If we suppose that the system is in the ground state, 
\begin{eqnarray} 
\Psi =\Psi _0=&\prod\limits_{j{\bf k}} {\left| {{{2\omega _{j{\bf k}}} \over {\pi  
\hbar  
}}} \right|^{{\textstyle{1 \over 4}}}\exp \left( {-\omega _{j{\bf k}}\hbar ^{- 
1}s_{j{\bf k}1}^2}  
\right)}\cr 
  &\times \prod\limits_{j^\prime {\bf k}^\prime } {\left| {{{2\omega _{j^\prime {\bf  
k}^\prime }} \over {\pi \hbar }}}  
\right|^{{\textstyle{1 \over 4}}}\exp \left( {-\omega _{j^\prime {\bf k}^\prime }\hbar  
^{-1}s_{j^\prime {\bf k}^\prime 2}^2}  
\right)} 
\end{eqnarray} 
then we can find the frictional forces by calculating the rate of excitation of the  
system. We do this in the first instance to second order perturbation theory to  
calculate the rate of excitation into the first excited state (one excitation per surface!), 
\begin{eqnarray} 
&\Psi _{J{\bf K}J^\prime {\bf K}^\prime }=&\cr 
&\left| {{{32\omega _{J{\bf K}}^3} \over {\pi \hbar ^3}}} \right|^{{\textstyle{1  
\over 4}}}s_{J{\bf K}1}^{}&\exp \left( {-\omega _{J{\bf K}}\hbar ^{-1}s_{J{\bf  
K}1}^2} \right)\cr 
&&\prod\limits_{j{\bf k}\ne J{\bf K}} {\left| {{{2\omega _{j{\bf k}}} \over {\pi  
\hbar }}} \right|^{{\textstyle{1 \over 4}}}\exp \left( {-\omega _{j{\bf k}}\hbar ^{- 
1}s_{j{\bf k}1}^2} \right)}\cr 
  \times &\left| {{{32\omega _{J^\prime {\bf K}^\prime }^3} \over {\pi \hbar ^3}}}  
\right|^{{\textstyle{1 \over 4}}}s_{J^\prime {\bf K}^\prime 2}^{}&\exp \left( {- 
\omega _{J^\prime {\bf K}^\prime }\hbar ^{-1}s_{J^\prime {\bf K}^\prime 2}^2}  
\right)\cr 
&&\prod\limits_{j^\prime {\bf k}^\prime \ne J^\prime {\bf K}^\prime } {\left|  
{{{2\omega _{j^\prime {\bf k}^\prime }} \over {\pi \hbar }}} \right|^{{\textstyle{1  
\over 4}}}\exp \left( {-\omega _{j^\prime {\bf k}^\prime }\hbar ^{-1}s_{j^\prime  
{\bf k}^\prime 2}^2} \right)} 
\end{eqnarray} 
The shift in the ground state energy can be written, 
\begin{eqnarray} 
\matrix{\Delta E&=&\sum\limits_{k_xk_y}^{} {{1 \over {16k^2\varepsilon  
_0^2}}}\exp \left( {-2kd} \right)\hfill\cr 
  &&\times \sum\limits_{JJ^\prime \;}^{} {\beta _{J{\bf K}}^2\beta _{J^\prime {\bf  
K}}^2\left| {{1 \over {16\,\omega _{J{\bf K}}\omega _{J^\prime {\bf K}}}}}  
\right|}{{2\left| {\omega _{J{\bf K}}+\omega _{J^\prime {\bf K}}} \right|} \over  
{\left| {\omega _{J{\bf K}}+\omega _{J^\prime {\bf K}}} \right|^2- 
k_x^2v^2}}\hfill\cr} 
\end{eqnarray} 
The imaginary part of $\Delta E$ gives the rate of excitation out of the ground state,  
therefore we can calculate the rate of working, 
\begin{eqnarray} 
\matrix{F_xv&=&{{dU} \over {dT}}=\sum\limits_{k_xk_y}^{} {{1 \over {2\pi  
^2}}}\exp \left( {-2kd} \right)\hfill\cr 
  &&\times \int_{-\infty }^{+\infty } {d\omega {\rm sgn} \left( \omega  \right)\,\Im  
\left[ {{{\varepsilon \left( \omega  \right)-1} \over {\varepsilon \left( \omega   
\right)+1}}} \right]}\int_{-\infty }^{+\infty } {d\omega^\prime {\rm sgn} \left(  
{\omega^\prime } \right)\;\Im \left[ {{{\varepsilon \left( {\omega^\prime } \right)-1}  
\over {\varepsilon \left( {\omega^\prime } \right)+1}}} \right]}\hfill\cr 
  &&\times \Im {{2\hbar \left| {\omega +\omega^\prime } \right|^2} \over {\left|  
{\omega +\omega^\prime } \right|^2-k_x^2v^2+i\eta }}\hfill\cr} 
\end{eqnarray} 
where we have substituted for $\beta _{J\bf K}^2$ the expression we calculated  
earlier. Performing the $\omega^\prime $ integration: 
\begin{equation} 
F_x={\hbar  \over \pi }\sum\limits_{k_xk_y}^{} {}\int_{-\infty }^{+\infty } {\exp  
\left( {-2kd} \right)k_x\,\Im \left[ {{{\varepsilon \left( {k_xv-\omega } \right)-1}  
\over {\varepsilon \left( {k_xv-\omega } \right)+1}}} \right]\Im \left[ {{{\varepsilon  
\left( \omega  \right)-1} \over {\varepsilon \left( \omega  \right)+1}}} \right]d\omega  
} 
\end{equation} 
We recognise in equation (48) the expression for the frictional force derived in  
equation (14) of section 2. 
 
\section{The Nature of Quantum Friction} 
First some general observations about our formula for friction. We note that there  
must be loss process operating in both surfaces. Note also the similarity of our  
formula for friction to the expression derived earlier for the Van der Waals force.  
However the significant difference in the frictional force is that the imaginary part of  
the response is used.  
 
We can make some statements about the inter relationship of distance and velocity  
dependence of friction. Note that by substituting, 
\begin{equation} 
{\bf k}={\bf k}v 
\end{equation} 
our formula (20) can be rewritten, 
\begin{eqnarray} 
\matrix{F_x&=&4\hbar {{L^2} \over {\left( {2\pi } \right)^3}}{1 \over  
{d^3}}{{d^3} \over {v^3}}\int_0^{+\infty } {k_x^\prime dk_x^\prime }\int_{-\infty  
}^{+\infty } {\exp \left( {-2k^\prime dv^{-1}} \right)dk_y^\prime }\hfill\cr 
  &&\times \int_0^{k_x^\prime} {\Im R_{2pp}^{}\left( \omega  \right)\Im  
R_{1pp}^{}\left( {k^\prime _x-\omega } \right)d\omega }\hfill\cr} 
\end{eqnarray} 
hence, 
\begin{equation} 
F_x={1 \over {d^3}}g\left( {{v \over d}} \right) 
\end{equation} 
where $g$ is an arbitrary function. If we assume that the particular form of the  
dielectric function results in a power law dependence on the velocity it follows that, 
\begin{equation} 
F\propto {{v^\mu } \over {d^{\mu +3}}} 
\end{equation} 
Therefore the $v-$dependence and $\mu-$dependence are inter linked. This may be a  
way of identifying these contributions to friction. 
 
Now let us explore what happens with various standard forms of the dielectric  
function. 
\subsection{constant $\varepsilon \left( \omega  \right)$} 
This is the simplest case, but possibly the least physical, and results in a frictional  
force of the form, 
\begin{equation} 
F=\left[ {\Im {{\varepsilon -1} \over {\varepsilon +1}}} \right]^2{{3\hbar v} \over  
{2^6\pi ^2d^4}} 
\end{equation} 
Note that our `rule of powers' is obeyed and that friction has the form expected if the  
vacuum behaved as a viscous fluid. The force falls rapidly as $d^4$. Nevertheless  
there is no exponential decay of the force because interactions between the two  
surfaces are conveyed by a massless particle, the photon. Note also the central role  
played by, 
\begin{equation} 
\Im {{\varepsilon -1} \over {\varepsilon +1}} 
\end{equation} 
which represents the reflection coefficient of the surface to $p-$polarised radiation,  
and is also proportional to the density of electromagnetic states immediately outside a  
surface. 
\subsection{constant conductivity} 
In this case the dielectric function has the form, 
\begin{equation} 
\varepsilon =1+i{\sigma  \over {\omega \varepsilon _0}} 
\end{equation} 
where $\sigma$ is the conductivity. For this case the integrals are tricky to evaluate  
except in the limiting cases of high and low velocities. However we sketch the  
qualitative form of the force below, and quote the limiting cases, 
\begin{equation} 
\matrix{F={{5\,\hbar \varepsilon _0^2v^3} \over {2^8\pi ^2\sigma ^2}}{1 \over  
{d^6}},\;\;\;\;\;\;\;\;\;\;\;\;\;\;\;\;\;\;v<<{{d\sigma } \over {\varepsilon _0}}\hfill\cr 
  F={{\hbar \sigma ^2} \over {32d^2\pi ^2v\varepsilon _0^2}}\ln \left(  
{{{v\varepsilon _0} \over {2d\sigma }}} \right),\;\;\;\;\;v>>{{d\sigma } \over  
{\varepsilon _0}}\hfill\cr} 
\end{equation} 
Since this is a more realistic instance, it is worth evaluating the force. We pause a  
moment to maximise the effect which we achieve approximately by maximising, 
\begin{equation} 
\Im {{\varepsilon -1} \over {\varepsilon +1}}=\Im {{i{\sigma  \over {\omega  
\varepsilon _0}}} \over {2+i{\sigma  \over {\omega \varepsilon _0}}}}={{{{2\sigma  
} \over {\omega \varepsilon _0}}} \over {4+\left( {{\sigma  \over {\omega  
\varepsilon _0}}} \right)^2}} 
\end{equation} 
i.e. we choose, 
\begin{equation} 
\sigma =2\omega \varepsilon _0\approx 2kv\varepsilon _0\approx 2d^{- 
1}v\varepsilon _0 
\end{equation} 
where we have recognised that the critical frequencies will be the highest for which  
loss occurs. Assuming a modest shear velocity and surfaces in atomic contact gives, 
\begin{equation} 
\matrix{d=10^{-10}\rm m\hfill\cr 
  v=1.0\ {\rm ms}^{-1}\hfill\cr} 
\end{equation} 
where we have recognised that the dielectric response of the surface will cut off at  
something like a screening length, typically $10^{-10}\rm m$ in a metal. Under these  
conditions we choose, 
\begin{equation} 
\sigma =0.1\;\left( {\rm ohm-m} \right)^{-1} 
\end{equation} 
which puts us more or less at the cross-over point of the two limiting formula, and at  
the maximum of the friction/velocity curve in figure 4. This sort of conductivity is not  
untypical of semi-metals such as carbon. We make a rough estimate of the frictional  
force by substituting into the high velocity formula (the least sensitive to the velocity)  
and find, 
\begin{equation} 
F\approx 3\times 10^3{\rm Nm}^{-2} 
\end{equation} 
However this result needs to be qualified: no two surfaces placed in contact will  
actually touch over their entire area. Estimates of the fractional area in contact vary  
around 0.001. Therefore if we assume that our surfaces have only this fractional  
intimate contact the observed force will be much smaller, 
\begin{equation} 
F_{\rm obs}\approx 3{\rm Nm}^{-2} 
\end{equation} 
In other words for a restricted class of materials, the semi metals, in which the  
electromagnetic density of states outside the surface is maximised in the relevant  
frequency range, this contribution to frictional forces is substantial. Furthermore,  
because of the power law dependence on the separation of surfaces, it will dominate  
the long range contributions as the other contribution, due to exchange of electrons,  
must always have an exponential decay in consequence of the finite work function of  
the electron. 
\subsection{frictional forces independent of velocity} 
The dielectric functions discussed above all lead to frictional forces {\it dependent} on  
the velocity, but experiment mainly measures forces {\it independent} of velocity. It  
is interesting to speculate on whether our model of photon exchange can reproduce  
velocity independence and under what circumstances.  
 
Inspecting equation (20) for the force we can see that the velocity dependence would  
be eliminated if the reflection coefficients had the form shown in figure 5. Under these  
circumstances the frequency integrand in (20) is sketched in figure 5. If we assume  
that the peaks dominate the integration, then the result will be independent of velocity.  
Of course this assumes a finite velocity such that the peaks are well separated. 
 
We might further speculate on the nature of a surface with a capacity to absorb low  
frequency radiation with large components of momentum parallel to the surface.  
Surface scientists are fond of invoking a `dirty' surface and such an object exactly fills  
the bill in this case. A surface on which there is a random assortment of loose massive  
fragments will absorb a great deal of momentum for little energy input. The fragments  
need only be massive relative to the fundamental particles, and nanometre sized lumps  
would be perfectly adequate. Figure 6 shows our model of a dirty surface, and we  
suggest that the electromagnetic reflection coefficient would show the low frequency  
peak sketched in figure 5. The width of the low frequency peak could then be  
predicted in terms of the mass, $M$, of the fragments: 
\begin{equation} 
\Delta \approx {{\hbar k_{\max }^2} \over M} 
\end{equation} 
Interestingly enough at low velocities the model predicts that the frictional force will  
rise rapidly when the peaks overlap. This will happen when, 
\begin{equation} 
k_xv\approx \Delta  
\end{equation} 
where   is the width of the low frequency peak. The effect is shown in figure 7. The  
critical velocity at which sliding friction turns over into static friction is predicted by  
the width, $\Delta $, in the low frequency peak in reflectivity: 
\begin{equation} 
k_{\max }v_c\approx \Delta  
\end{equation} 
where $k_{\max }$ is the largest wave vector that can be excited in the system, which  
we expect to be of the order of $10^{+11}{\rm m}^{-1}$. 
 
\section{Other Contributions to Friction} 
Our perspective is that friction arises through exchange of momentum carrying  
particles between surfaces. We have considered the photon, but other exchanges are  
possible: atoms may be exchanged as happens in the case of friction with wear  
[21,22], but the most obvious competitor to the photon is the electron. Note that the  
{\it phonon}, as opposed to the {\it photon}, is not a particle that can have any  
existence separate from the solid, it does not exist in vacuum and therefore is not on  
the list of exchangeable particles though phonon exchange can be mediated by a  
another particle. 
 
The interaction of two surfaces through exchange of electrons is a vast subject and  
covers nearly the whole of chemical bonding at surfaces. Therefore, since we are only  
interested in a simple comparison with the photon case, we choose the most  
elementary possible model capable of generating friction within the context of two  
smooth translationally invariant surfaces. We shall assume the surfaces to be made of  
the same material, and the electrons to be defined by a spherical Fermi surface. 
 
When the two surfaces are at rest relative to one another, there is no exchange of  
electrons because the exclusion principle forbids tunnelling into filled states. When  
the surfaces are in relative motion a small slit of states appears on either side of the  
Fermi surfaces, where tunnelling is allowed, see figure 8. We can work out the rate at  
which the surfaces exchange momentum through this tunnelling mechanism: 
\begin{equation} 
F_x=2\int_0^\pi  {\hbar k_F^4v{{L^2} \over {\left( {2\pi } \right)^3}}\cos ^2\left(  
\theta  \right)\sin ^2\left( \theta  \right)}\exp \left[ {-2d\sqrt {{{2m\phi } \over {\hbar  
^2}}+k_F^2\cos ^2\theta }} \right]d\theta  
\end{equation} 
where $d$ is the separation between the two surfaces, $\phi $ is the work function  
relative to vacuum, $k_F$ is the Fermi momentum, $m$ is the electronic mass, $v$ is  
the shear velocity, and $\theta$ is the angle between the electron momentum and $\bf  
v$. 
 
In the limit that the surfaces are in contact, 
\begin{eqnarray} 
\matrix{F_x&=&2\int_0^\pi  {\hbar k_F^4v{{L^2} \over {\left( {2\pi }  
\right)^3}}\cos ^2\left( \theta  \right)\sin ^2\left( \theta  \right)}d\theta \hfill\cr 
  &=&{1 \over 2}\int_0^\pi  {\hbar k_F^4v{{L^2} \over {\left( {2\pi } \right)^3}}\sin  
^2\left( {2\theta } \right)}d\theta \hfill\cr 
  &=&{\pi  \over 4}\hbar k_F^4v{{L^2} \over {\left( {2\pi } \right)^3}}={{L^2}  
\over {32\pi ^2}}\hbar k_F^4v\hfill\cr} 
\end{eqnarray} 
Substituting a Fermi momentum typical of aluminium and a typical shear velocity, 
\begin{eqnarray} 
\matrix{k_F&=&1\ {\rm au}\approx 10^{+10}\rm m^{-1},\hfill\cr 
  v&=&1\ {\rm ms}^{-1}\hfill\cr} 
\end{eqnarray} 
the force is, 
\begin{equation} 
L^{-2}F_x={1 \over {32\pi ^2}}\hbar k_F^4v={{10^{-34}\times 10^{+40}\times 1}  
\over {32\pi ^2}}=3.16\times 10^{+3}{\rm Nm}^{-2} 
\end{equation} 
Note that this electronic contribution to friction is comparable in magnitude to that  
obtained from the tunnelling of photons: compare equation (61). 
 
The main difference between photonic and electronic friction is the dependence on  
separation, $d$, between the surfaces. In the photonic case there is nearly always a  
power law whereas in the electronic case the existence of a finite work function  
dictates that the force decays exponentially with $d$. Photons always dominate at  
large distances. 
 
\section{Can Sheared Interfaces Emit Light?} 
The friction we have discussed so far involves frequencies typically in the GHz range,  
much lower than optical frequencies. The frequency is limited by $k$,  the wave  
vector parallel to the surface of the radiation, and the relative velocity of the surfaces, 
\begin{equation} 
\omega \le kv 
\end{equation} 
as we have already discussed. Since the screening length in a material is of the order  
of the separation between electrons, the highest frequencies possible at shear  
velocities of the order of $1{\rm ms}^{-1}$ are no more than a few GHz. Therefore  
there is no emission of visible light by this mechanism unless the shear velocity is  
impossibly large. 
 
GHz microwaves can be produced, but to be observed must be ejected into a  
transparent dielectric and have a real wave vector: 
\begin{equation} 
K_z^\prime =\sqrt {\varepsilon \,\omega ^2c_0^{-2}-k_x^2-k_y^2+i\delta } 
\end{equation} 
where the prime denotes that we are in the dielectric. For $K_z^\prime$ to be real, 
\begin{equation} 
\varepsilon \,\omega ^2c_0^{-2}\approx \varepsilon \,k^2v^2c_0^{-2}>k_x^2+k_y^2 
\end{equation} 
which requires, 
\begin{equation} 
\varepsilon \,>\left( {{{c_0^{}} \over v}} \right)^2 
\end{equation} 
Our conclusion is that, unless we assume improbably large velocities or huge values  
of $\epsilon$, no free radiation can be emitted from smooth sheared surfaces. The  
situation is different if we allow the surfaces to be rough, but that is beyond the scope  
of this paper. 
 
\section{Conclusions} 
We have shown that friction exists between sheared smooth dielectric surfaces at  
$T=0$, provided that we take quantum fluctuations into account. If the surfaces are  
almost in physical contact, the frictional forces may be comparable to other  
contributions observed in everyday situations, provided that the surfaces have a high  
density of electromagnetic states in the GHz region: in other words resistivity of the  
order of  $1{\rm m\Omega}$. Details of how friction depends on surface separation  
and velocity vary with the materials, but in general friction decays with a power law  
dependence on surface separation, in a manner linked to the velocity dependence and  
can be expected to be the dominant frictional force at large separations, just as the  
Van der Waals force dominates in this regime. 
 
Although finite temperatures are beyond the scope of this paper, we can expect some  
modifications to our conclusions at room temperature because the excitations created  
by friction are of the order of $k_BT_{room}$. 
 
The formula for quantum friction (20) is very simple and given a few assumptions can  
be derived in a few lines of algebra alongside the well known formula for the  
attractive force between surfaces. A more careful quantum treatment associates  
quantum friction with emission of pairs of correlated photons, one into each surface,  
but gives the same formula as the rough approach.  
 
Brief consideration was given to friction arising from exchange of electrons: a  
different mechanism from the electron-hole pair creation considered previously (the  
latter is in reality a photon exchange process). Tunnelling of electrons between  
surfaces can also give a substantial frictional force, but exponential decay of the  
electron wave functions means that these forces are shorter ranged that photon based  
forces. 
 
Finally we asked whether free light could be emitted from a sheared interface. It can  
but under extreme conditions of shear velocity, or of material properties that render it  
practically impossible. This conclusion is reached within the context of smooth  
surfaces. If the surfaces are rough on an atomic scale momentum conservation  
arguments lying behind our conclusions are no longer valid. Rough surfaces may emit  
light on being sheared. 
 
\ack 
I thank Rufus Ritchie for many helpful references which guided my thinking on this  
problem, Erio Tosatti for drawing my attention to his book, such a valuable source of  
further references, and Pedro Echenique for his insistence that life must be simpler  
than in sections 3 and 4.


\begin{figure}
\epsffile{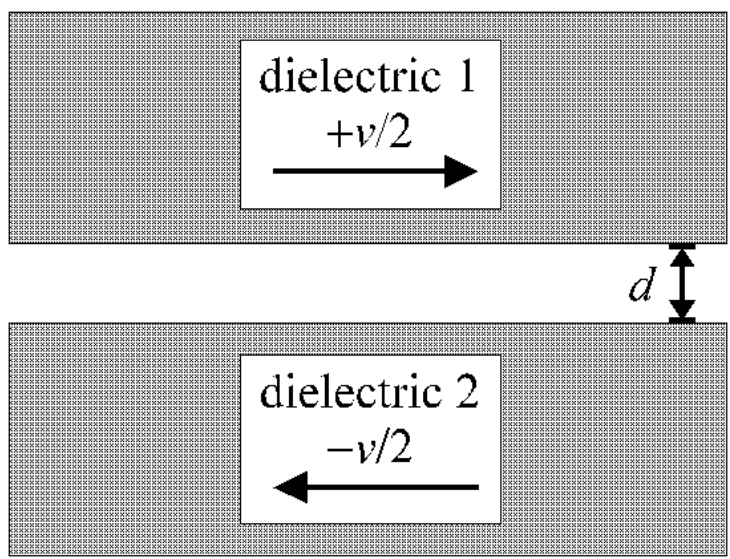}
\caption{Perfectly smooth dielectric surfaces shear against one another with relative  
velocity $v$. Is there a frictional force?}
\label{fig1}
\end{figure}

\begin{figure}
\epsffile{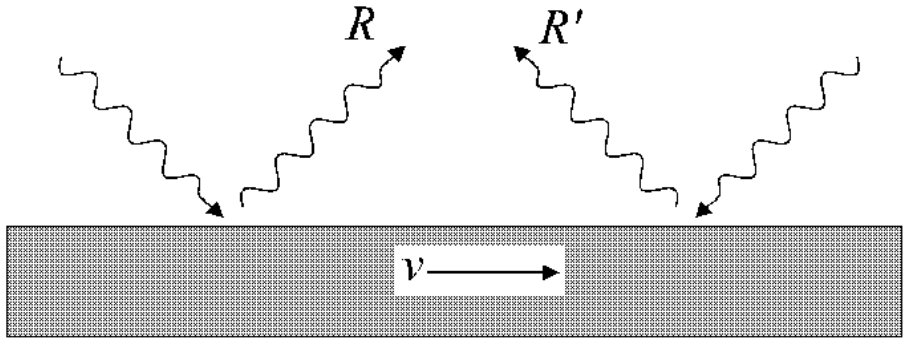}
\caption{We can detect the motion of a smooth dielectric surface by measuring its  
reflection coefficient with and against the motion as shown in the figure. In the  
reference frame of the dielectric the two incident waves are Doppler shifted in  
opposite senses and therefore reflect differently from the surface.}
\label{fig2}
\end{figure}

\begin{figure}
\epsffile{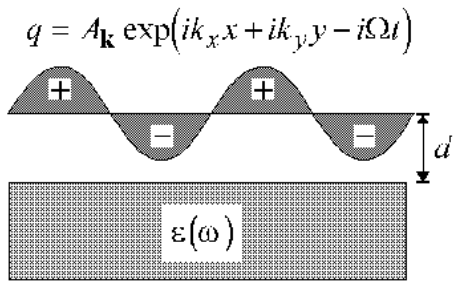}
\caption{A sheet of charge sits a distance $d$ above a dielectric surface.}
\label{fig3}
\end{figure}

\begin{figure}
\epsffile{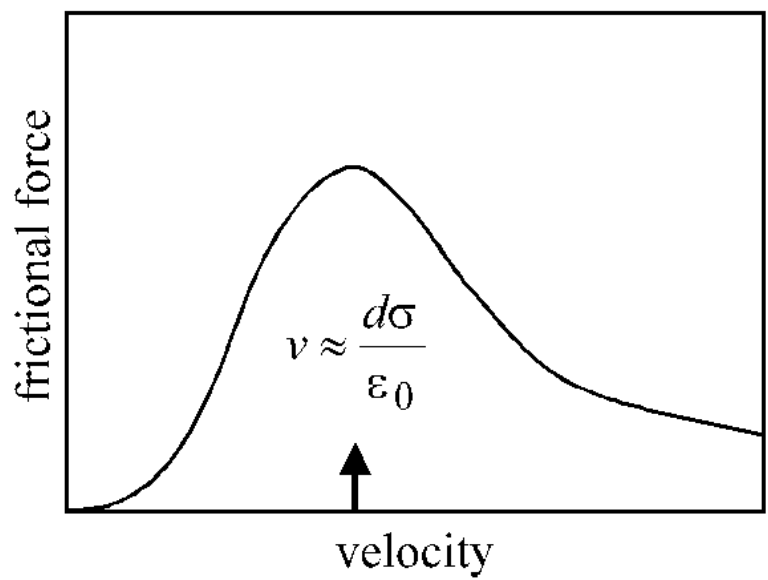}
\caption{The frictional force between two conducting surfaces separated by distance  
$d$. At low velocities the forces increase as $v^3$, at high velocities it decreases as  
$v^{-1}\ln v$, reaching a maximum at $v\approx d\sigma \epsilon ^{-1}_0$.}
\label{fig4}
\end{figure}

\begin{figure}
\epsffile{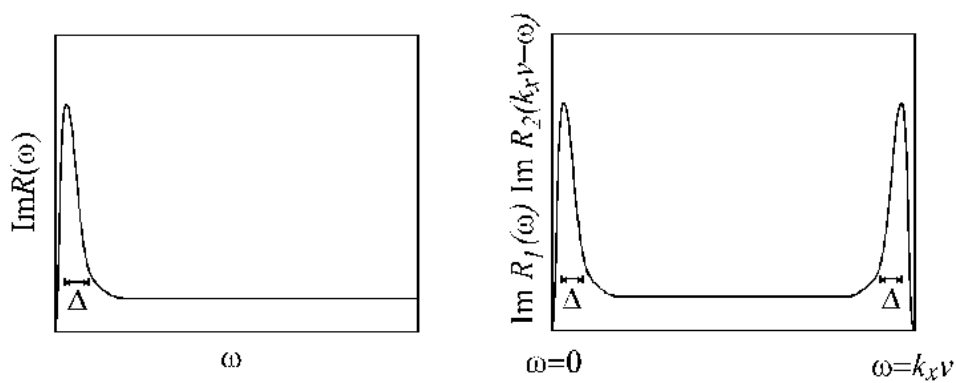}
\caption{The reflection coefficient shown on the left implies that the surface can  
sustain many low frequency excitations which absorb momentum without absorbing  
much energy. At finite frequencies the coefficient is nearly constant. On the right is  
shown the corresponding the frequency integrand in equation (20): if the peaks are  
strong enough the integral will be independent of $v$.}
\label{fig5}
\end{figure}

\begin{figure}
\epsffile{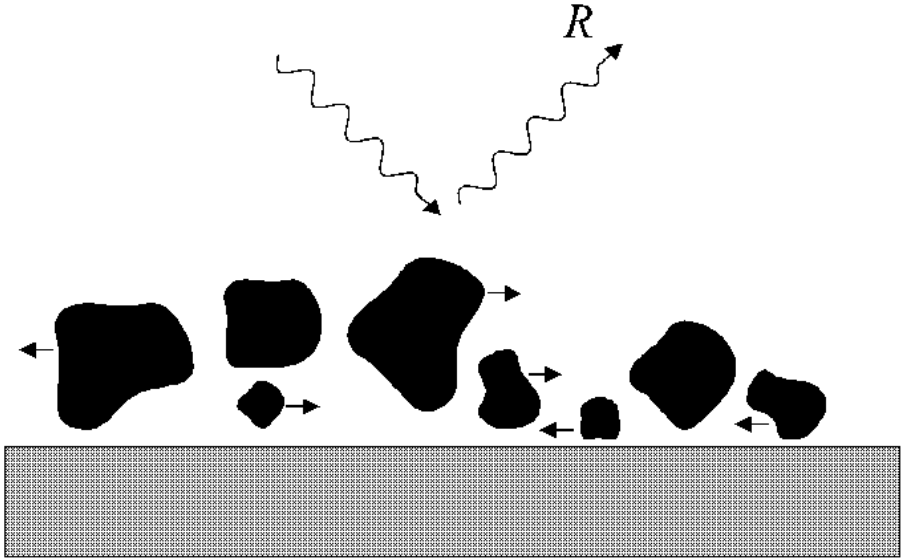}
\caption{Model of a `dirty' surface: low frequency excitations of massive particles on  
the surface absorb plenty of momentum but little energy leading to a low frequency  
peak in absorption of incident radiation.}
\label{fig6}
\end{figure}

\begin{figure}
\epsffile{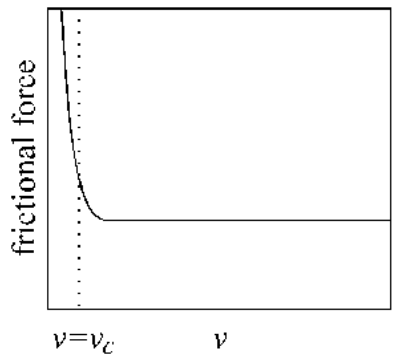}
\caption{The frictional force as a function of velocity as predicted by our model of a  
`dirty' surface.}
\label{fig7}
\end{figure}

\begin{figure}
\epsffile{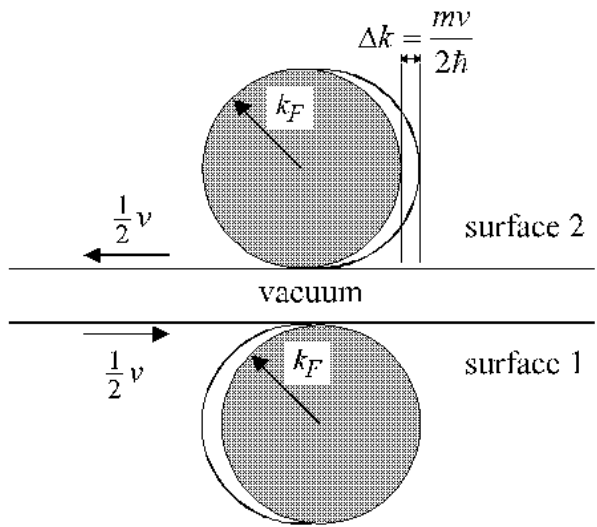}
\caption{When two surfaces move relative to one another, the Fermi surfaces shift  
slightly in the direction of movement, creating a thin annulus of states into which  
electrons can tunnel from the other surface. This creates a momentum exchange and  
hence a force between the surfaces.}
\label{fig8}
\end{figure}

\end{document}